\documentclass[%
 aip,
 amsmath,amssymb,
 reprint,%
]{revtex4-1}

\usepackage{graphicx}
\usepackage{dcolumn}
\usepackage{bm}

\usepackage[utf8]{inputenc}
\usepackage[T1]{fontenc}
\usepackage{mathptmx}
\usepackage{etoolbox}

\usepackage{multirow}

\makeatletter
\def\@email#1#2{%
 \endgroup
 \patchcmd{\titleblock@produce}
  {\frontmatter@RRAPformat}
  {\frontmatter@RRAPformat{\produce@RRAP{*#1\href{mailto:#2}{#2}}}\frontmatter@RRAPformat}
  {}{}
}%
\makeatother
\begin{document}

\preprint{AIP/123-QED}

\title[Simultaneous measurement of extensional stress and flow birefringence field for uniaxially extending worm-like micellar solutions]{Simultaneous measurement of extensional stress and flow birefringence field for uniaxially extending worm-like micellar solutions}

\author{M. Muto}%
 \email{muto.masakazu@nitech.ac.jp and tamano.shinji@nitech.ac.jp}
\affiliation{ 
Department of Mechanical Engineering, Nagoya Institute of Technology, Gokiso, Showa-ku, Nagoya-shi, Aichi, Japan
}%
\author{T. Yoshino}%
\affiliation{ 
Department of Mechanical Engineering, Nagoya Institute of Technology, Gokiso, Showa-ku, Nagoya-shi, Aichi, Japan
}%
\author{S. Tamano}%
\affiliation{ 
Department of Mechanical Engineering, Nagoya Institute of Technology, Gokiso, Showa-ku, Nagoya-shi, Aichi, Japan
}%

\date{\today}

\begin{abstract}
The present study proposes a novel and simple rheo-optical technique to investigate the relation between the rheology of complex fluids and their internal structural deformation under uniaxial extensional flow.
The macroscale results of viscoelasticity from rheological measurements and microscale results of birefringence from optical measurements are combined to evaluate the microstructural deformation and orientation state inside the fluids under extensional stress. 
The proposed technique combines a liquid dripping method with a high-speed polarization camera to measure the extensional stress and flow-induced birefringence field simultaneously.
In the liquid dripping method, temporal evolution images of the liquid filament diameter for fluids dripping from a nozzle are measured to obtain the extensional stress loading on the liquid filament.
These images are captured with a high-speed polarization camera connected to a micro polarization element alley, enabling high-speed imaging of the birefringent field.
Worm-like micellar solutions of cetyltrimethylammonium bromide (CTAB) and sodium salicylate (NaSal) with varying concentrations of CTAB and NaSal are employed as the measurement targets. 
Consequently, we successfully visualized temporally developing images of the birefringence field of uniaxially extending worm-like micellar solutions induced by the orientation of micelles toward the extensional direction.
Furthermore, the proposed technique supports investigating the conditions for establishing the stress-optical rule, which is the linear relation between stress and birefringence for complex fluids.
The stress-optical coefficient, a proportionality constant indicating the sensitivity of birefringence to stress, is analyzed from these measurements.
The stress-optical coefficient under uniaxial extensional flow is confirmed to be comparable to that under shear flow.
\end{abstract}

\maketitle

\section{\label{introduction}Introduction \protect}
To investigate the relation between the rheology of complex fluids and their internal structural deformation under stress loading, a rheo-optical technique \cite{noto2020applicability, philippoff1957flow, lodge1955variation, lodge1956network, gortemaker1976re, yevlampieva1999flow, meyer1993investigation, chow1984response, ito2015temporal, ito2016shear, shikata1994rheo, wheeler1996structure, C_Takahashi, kadar2021cellulose, muto2022unsteady, haward2011extensional, haward2012extensional, janeschitz2012polymer, doyle1998relaxation,humbert1998flow, pathak2006rheo,ge2012rheo,decruppe2003flow,humbert1998stress,sridhar2000birefringence,talbott1979streaming} involving simultaneous rheological and optical measurements is used. 
One of the optical measurements used in the rheo-optical technique is birefringence measurement \cite{bass2010handbook,rastogi2015digital}. 
Quantifying the changes in the optical properties of a component using birefringence measurements enables estimating the orientation state of the microstructure inside complex fluids. 
By combining macro-and micro-scale results from viscoelasticity and birefringence measurements, respectively, the structural deformation inside complex fluids under stress loading can be comprehensively evaluated. 
The impact of this method is derived by implementing the quality of optical products, such as pickup lenses, optical discs, and liquid crystal films. 
For example, in the injection molding method \cite{mckelvey1962polymer,angstadt2006investigation, kim2005modeling, min2011experimental, lin2019grey} for manufacturing optical products, the molten material is exposed to strong shear and extensional flows during the injection molding, inducing residual stress. 
Because birefringence is proportional to stress, residual stress causes birefringence, which influences the quality of optical products, such as optical aberrations and color irregularities \cite{lin2019grey}. 
Therefore, product qualities can be improved by understanding the stress loading state during injection molding and controlling the induced birefringence.

The proportional relation between the birefringence and stress is known as the stress-optical rule \cite{rastogi2003photomechanics,rastogi2013optical,noto2020applicability,janeschitz2012polymer}. 
Applying stress to the component that exhibits birefringence, the birefringence $\delta n$ for the principal stress difference $\sigma_{\parallel}-\sigma_{\perp}$ is expressed as 
\begin{equation}
\delta n = n_{\parallel}-n_{\perp} = C (\sigma_{\parallel}-\sigma_{\perp})
\label{eq1},
\end{equation}
where $n_{\parallel}$ and $n_{\perp}$ are the refractive indices for polarizations parallel and perpendicular to the oriented birefringent objects, such as polymer chains and worm-like micelles, respectively.
The principal stress difference is the difference between the principal stresses perpendicular to each other. The principal stresses ($\sigma_{\parallel}$ and $\sigma_{\perp}$) are proportional to the refractive indices ($n_{\parallel}$ and $n_{\perp}$) corresponding to each principal stress direction.
The proportional coefficient $C$ is called the stress-optical coefficient and indicates the sensitivity of the birefringence of the component to stress.
A general birefringence measurement technique enables the optical measurement of phase retardation $\delta$ as the integrated value of birefringence $\delta n$, expressed as follows:
\begin{equation}
\delta = \int \delta n\,dz = Cl(\sigma_{\parallel}-\sigma_{\perp})
\label{eq2},
\end{equation}
where the $z$-axis direction is the direction of light travel and $l$ is the optical path length. 
The measured data of the phase retardation $\delta$ and stress-optical coefficient $C$ of the component must be obtained to investigate the stress applied inside the component.

\begin{figure}[b!]
\includegraphics[width=\columnwidth]{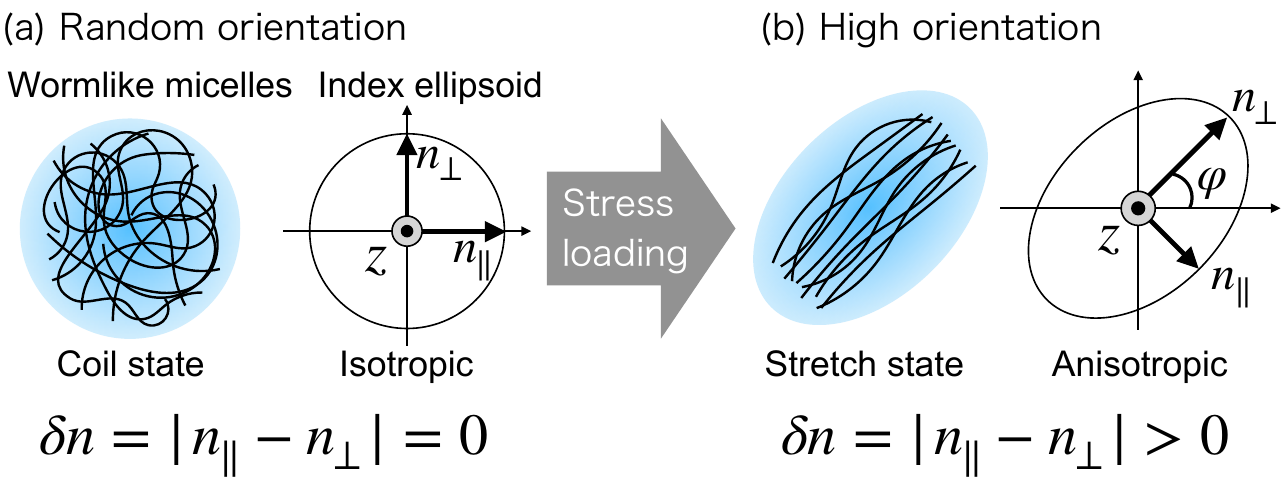}
\caption{\label{fig1} 
Birefringence induced by a change in the orientation state of worm-like micelles under stress loading. Without stress loading, the worm-like micelles in the solutions have random orientations in the coil state; thus, flow-induced birefringence does not occur. When applying stress to the solutions, the worm-like micelles extend, tilt at the orientation angle $\varphi$, and become highly oriented, increasing birefringence $\delta n$. }
\end{figure}

It is necessary to demonstrate the intrinsic physical property of the stress-optical coefficient to validate the stress-optical rule.
For solids, several studies have investigated\cite{rastogi2013optical,rastogi2003photomechanics,ebisawa2007mechanical,sun2023measurement} obtaining the stress-optical coefficient using the photoelastic method. 
For fuluids, on the other hand, some studies have investigated flow birefringence measurement of complex fluids \cite{kadar2021cellulose,muto2022unsteady,lane2022birefringent,haward2019flow,haward2011extensional,haward2012extensional,iwata2019local,zhao2016flow,janeschitz2012polymer,doyle1998relaxation,cardiel2016formation,sridhar2000birefringence,talbott1979streaming}, and measured the stress-optical coefficients under shear flow \cite{shikata1994rheo,wheeler1996structure,C_Takahashi,ito2016shear,haward2011extensional,humbert1998flow,ge2012rheo,decruppe2003flow,humbert1998stress}. 
Among the viscoelastic fluids that induce birefringence, worm-like micellar solutions are typical examples, and their birefringence fields have been widely measured \cite{shikata1994rheo,wheeler1996structure,C_Takahashi,ito2015temporal,ito2016shear,haward2019flow,cardiel2016formation,ge2012rheo,decruppe2003flow,humbert1998stress}. 
Shikata et al. \cite{shikata1987micelle,shikata1988micelle,shikata1989micelle,shikata1988nonlinear} reported that worm-like micellar solutions of cetyltrimethylammonium bromide (CTAB) and sodium salicylate (NaSal) form extremely long worm-like micelles at relatively low concentrations and exhibit remarkable viscoelasticity.
In addition, CTAB/NaSal solutions exhibit birefringence under stress loading and are often used in birefringence measurements.
To measure the stress-optical coefficient of CTAB/NaSal solutions under shear flow, Shikata et al. \cite{shikata1994rheo}, Wheeler et al. \cite{wheeler1996structure}, Takahashi et al. \cite{C_Takahashi}, and Humbert et al. \cite{humbert1998flow} reported using a rheo-optical techinque with a coaxial cylinder rheometer or a parallel plate rheometer.
Ito et al. \cite{ito2015temporal,ito2016shear} confirmed that the birefringence of worm-like micellar solutions under shear flow induced by a coaxial cylinder rheometer is synchronized with its stress fluctuation and generates a shear-induced structure.
The birefringence of worm-like micelles can be explained by the optical anisotropy induced by the index ellipsoid deformation, as shown in Fig. \ref{fig1}. 

Although flow birefringence measurements under extensional and shear flows are important for understanding the structural deformation of complex fluids, only a few studies have explored the phenomenon under extensional  flow\cite{pathak2006rheo,haward2011extensional,rothstein2002comparison,sridhar2000birefringence,talbott1979streaming}. 
Pathak and Hudson \cite{pathak2006rheo} confirmed the birefringence of worm-like micellar solutions under extensional flow using a cross-slot channel.
Haward et al. \cite{haward2011extensional} also used cross-slot channels to confirm the orientation of atactic polystyrene under extensional flow. 
Rothstein et al. \cite{rothstein2002comparison}, Sridhar et al. \cite{sridhar2000birefringence}, and Talbott et al. \cite{talbott1979streaming} measured the birefringence of uniaxially extended filaments of polystyrene solutions using a filament- stretching rheometer. 
However, to the best of the author's knowledge, there are no reports of measuring the stress-optical coefficients of uniaxially extending worm-like micellar solutions.

The present study developed  an unsteady and noncontact rheo-optical technique to measure the viscoelasticity and birefringence of uniaxially extending worm-like micellar solutions simultaneously.
This method that combines the liquid dripping method \cite{muto2023rheological,tamano2017dynamics,amarouchene2001inhibition} with a high-speed polarization camera enables visualization of the birefringence field of complex fluids under uniaxial extensional flow and obtainment of their stress-optical coefficients. 
In the liquid dripping method, the extensional stress affecting the liquid filament is obtained contactless by capturing the filament radius, which decays as it is uniaxially extended. 
Furthermore, the same derivation formulas as those in the dripping-on-to-substrate capillary breakup extensional rheometry (CaBER-DoS) method \cite{dinic2020flexibility, dinic2019macromolecular,dinic2017pinch,dinic2015extensional,sur2018drop,mckinley2005visco,mckinley2000extract,clasen2006dilute,omidvar2019detecting,wu2020effects} can be used to derive different viscoelastic data. 
By capturing temporal images of the viscoelastic behavior of complex fluids using a high-speed polarization camera, used to measure the birefringence field from the polarization information of the light transmitted through the liquid filament, the birefringence corresponding to the extensional stress can be simultaneously measured.
The uniaxial extensional flow allows the extensional direction of worm-like micellar solutions, and the orientation direction of the micelles is in the same direction, which has the following two advantages.

\begin{figure}[tb!]
\includegraphics[width=0.98\columnwidth]{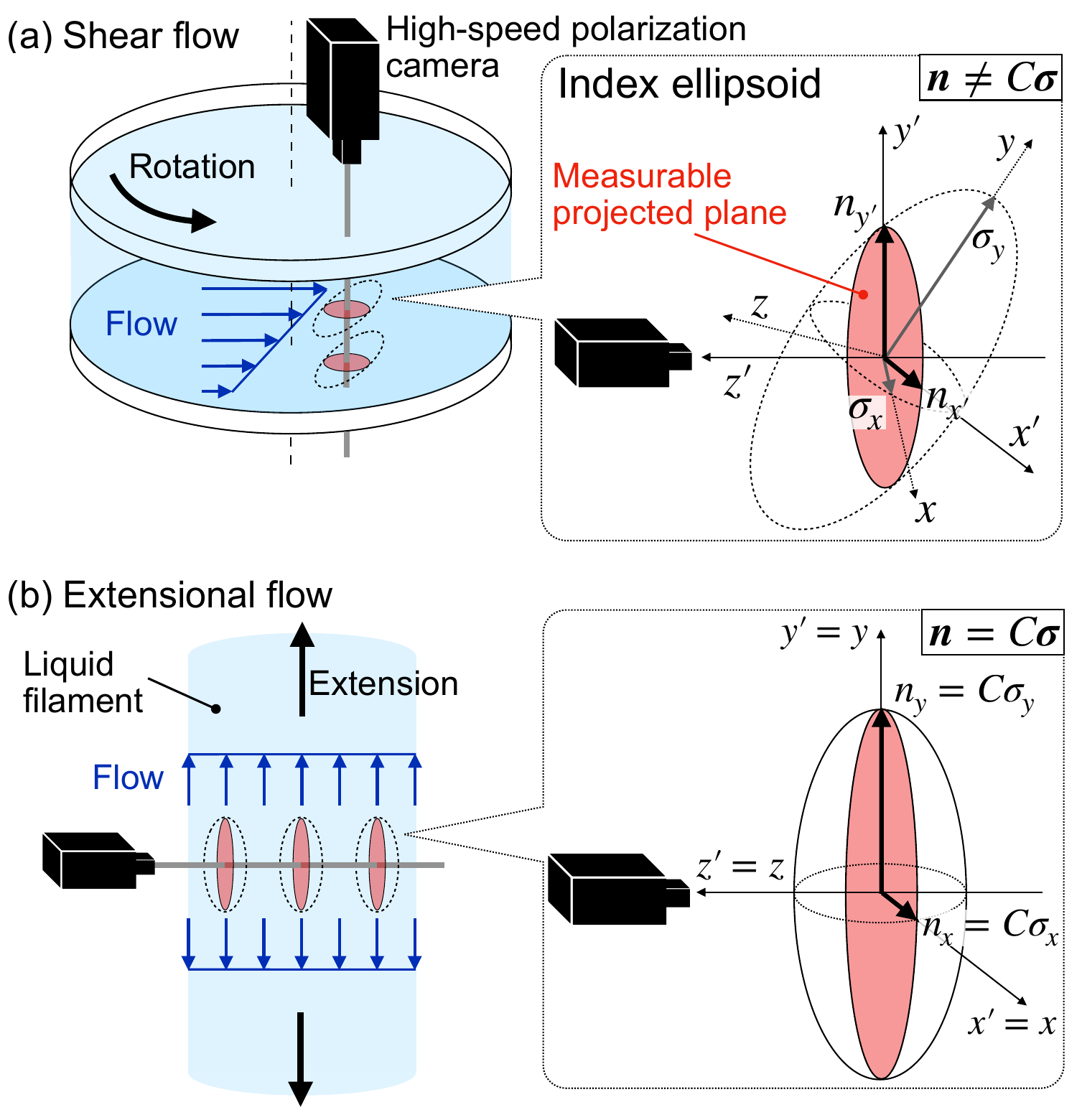}
\caption{\label{fig2} 
Conditions of index ellipsoids under (a) shear and (b) uniaxial extensional flows. 
Here, the major axis ($n_{\perp}$) and minor axis ($n_{\parallel}$) of refractive indices in the index ellipsoid correspond to $n_{y}$ and $n_{x}$, respectively. $x, y, z$ denote the coordinate system based on the refractive index ellipsoid, and $x', y', z'$ denote the coordinate system aligned with the optical axis of the high-speed polarization camera. In case (b), both coordinate systems coincide.}
\end{figure}

First, the major axis of the refractive index in the index ellipsoid is oriented in the extensional direction; thus, the stress-optical rule is valid. 
The birefringence measurement in the rheo-optical technique allows measuring the differences between the major axis ($n_{\perp}$) and minor axis ($n_{\parallel}$) of the refractive indices in the plane perpendicular to the optical axis. 
In the birefringence measurements under shear flow using a parallel plate (Fig. \ref{fig2}(a)), it is difficult to obtain the stress-optical coefficient. 
This is because the index ellipsoid is tilted toward the optical axis, and measuring the refractive index which corresponds to the tilted component of the principal stress vector ($\bm{\sigma}$) becomes challenging. 
Therefore, the stress-optical rule that indicates the relation between the refractive index vectors ($\bm{n}$) and principal stress vectors breaks down ($\bm{n} \neq C \bm{\sigma}$), resulting in inaccurate measurements of the birefringence corresponding to the principal stress desired for measurement under shear flow. 
However, birefringence measurements under uniaxial extensional flow using the liquid dripping method overcame this problem (Fig. \ref{fig2}(b)). 
The index ellipsoid oriented perpendicular to the optical axis has its major axis pointing in the extensional direction and is not tilted toward the optical axis. 
Therefore, the stress-optical rule is satisfied ($\bm{n} = C \bm{\sigma}$), enabling accurate measurement of the birefringence data.

Second, only the principal stress in one direction affects the uniaxially extended liquid filament, simplifying Eq. (\ref{eq2}) for the principal stress difference. 
As there is no flow in the radial direction of the filament, the principal stress in the horizontal component can be neglected ($\sigma_{\parallel} = 0$). 
Furthermore, the principal stress in the vertical component is rewritten as uniaxial extensional stress ($\sigma_{\perp} = \sigma_{\rm \hspace{1pt} E}$), which is calculated by measuring the filament radius. 
By observing the extension of the liquid filament in the lateral direction, the optical path length at which light travels inside the filament can be regarded as the droplet diameter ($l=2R$). 
Therefore, Eq. (\ref{eq2}) is rewritten as follows:
\begin{equation}
\delta = 2 C R \sigma_{\rm \hspace{1pt} E} 
\label{eq3}.
\end{equation}
Thus, the stress-optical coefficient is obtained by
\begin{equation}
C=\delta n / \sigma_{\rm \hspace{1pt} E}
\label{eq4}.
\end{equation}
This is valid only when the incident polarized light passes through the center of the filament.
The present study performed rheo-optical measurements of the birefringence field in a uniaxial extensional flow of worm-like micellar solutions: CTAB/NaSal solutions. 
The stress-optical coefficients of the worm-like micellar solutions under uniaxial extensional flow, calculated based on the birefringence and extensional stress data, were evaluated.

\section{\label{method}Methods \protect}
\begin{figure*}[tb!]
\includegraphics[width=0.74\textwidth]{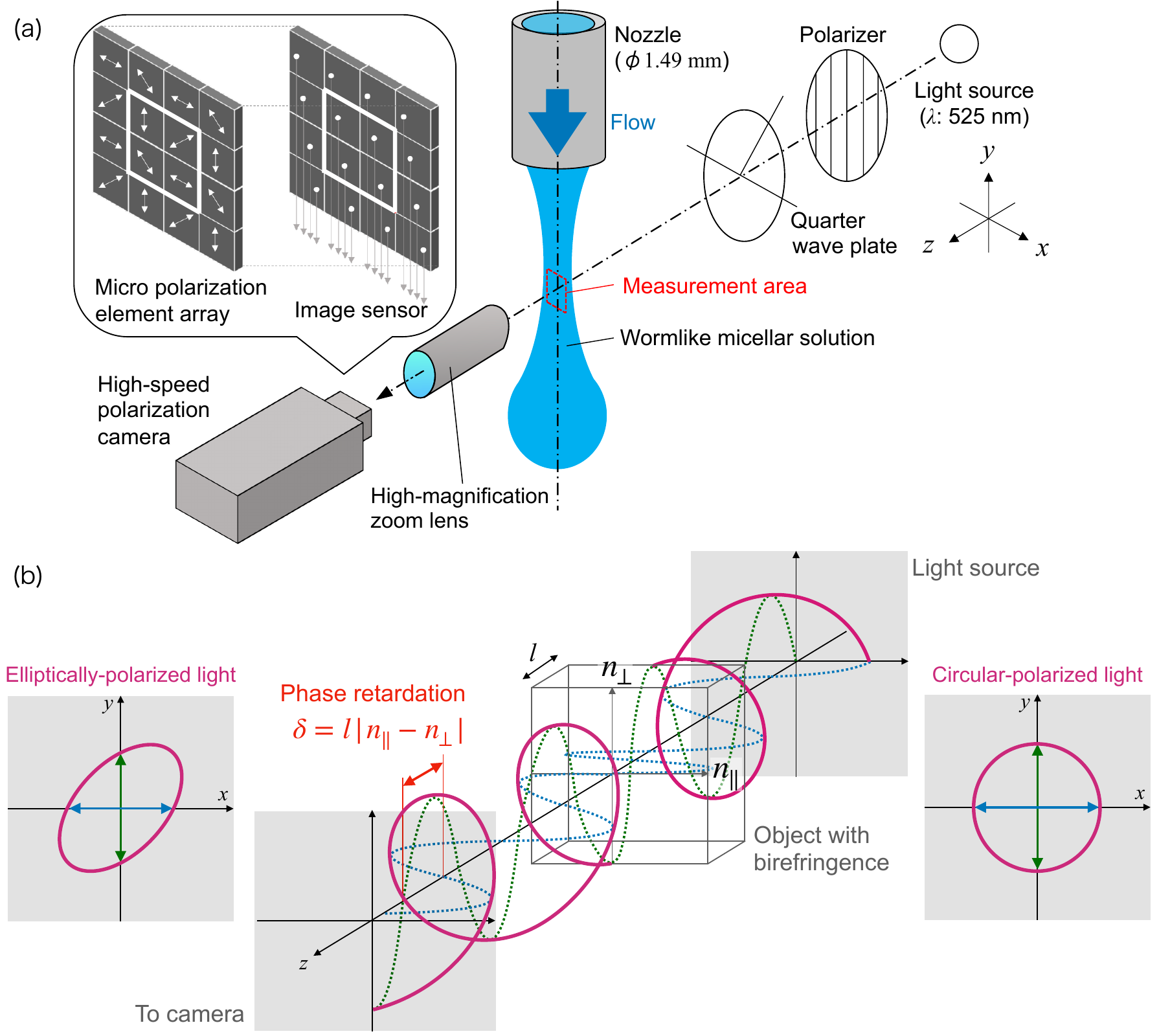}
\caption{\label{rheo_optical_technique}
Rheo-optical technique using a high-speed polarization camera and liquid dripping method. Schematics of (a) experimental setup and (b) measurement principle of the phase retardation induced by changes in the polarization state of a birefringent object (worm-like micellar solutions).}
\end{figure*}

We propose a simple calibration technique that combines the liquid-dripping method with a high-speed polarization camera (CRYSTA PI-1P, Photron Ltd.), as shown in Fig. \ref{rheo_optical_technique}(a). 
Our concept can be realized using a simple setup: a target sample (uniaxially extending worm-like micellar solution) between an LED light source (SOLIS-525C, Thorlabs, Inc.) and a polarization camera with a high-magnification zoom lens (Leica Z16APO, Leica Microsystems, Inc.).
This section describes the measurement principle of liquid dripping and rheo-optical techniques and the specifications of worm-like micellar solutions.

\subsection{\label{liquid dripping method}Liquid dripping method}
The experimental setup for the liquid-dripping method was based on that used in our previous study \cite{muto2023rheological}.
A fluid dispensing device with a syringe pump (Pump 11 Elite, Harvard Apparatus Ltd.) delivers a droplet of the worm-like micellar solution at a relatively low flow rate.
A liquid filament is formed as the droplet slowly drips from the nozzle with an outer diameter $R_0$ = 1.49 mm.
By measuring the minimum necking radius $R$ of the filament, which decreases as it extends, the extensional stress, viscosity, strain, strain rate, and relaxation time can be calculated \cite{muto2023rheological,tamano2017dynamics,amarouchene2001inhibition}.

\begin{figure}[t!]
\includegraphics[width=0.9\columnwidth]{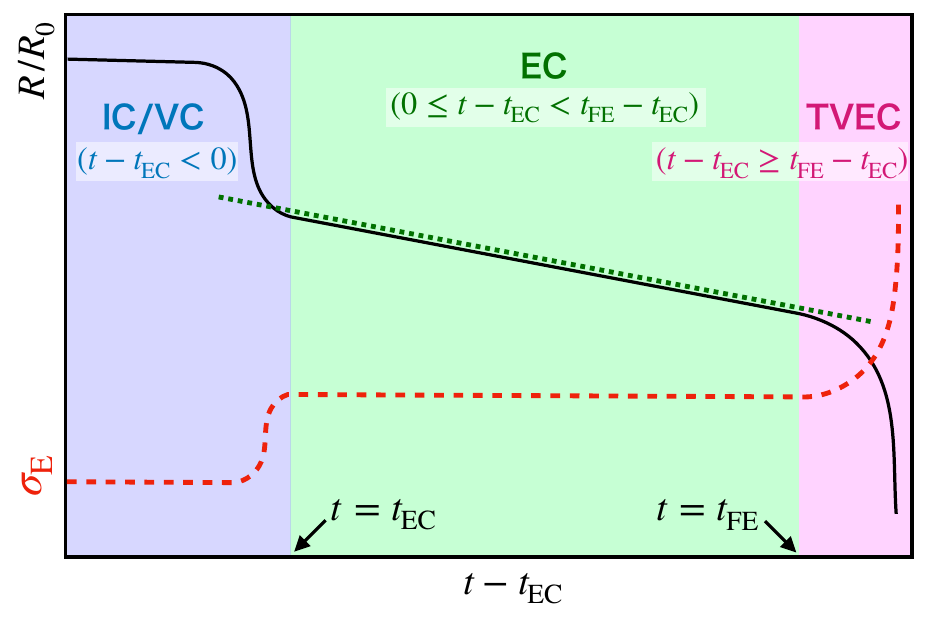}
\caption{Schematic of temporal evolution of radius ratio of liquid filament $R/R_0$ and extensional stress $\sigma_{\rm \hspace{1pt} E}$ for viscoelastic fluids.}
\label{EC_regime}
\end{figure}

To derive these physical quantities of viscoelasticity, the measurement data for the radius ratio $R/R_{0}$ of the liquid filament in the elasto-capillary (EC) regime must be extracted, as shown in Fig. \ref{EC_regime}.
Here, $R_{0}$ is the initial radius: the outer radius of the nozzle in the present study.
The horizontal axis of the graph is $t-t_{\mathrm{\hspace{1pt} EC}}$, where $t_{\mathrm{\hspace{1pt} EC}}$ denotes the start of the EC regime. 
The time evolution of the radius ratio $R/R_{0}$ is classified into three regimes \cite{dinic2019macromolecular}: inertio-capillary/visco-capillary (IC/VC), EC, and terminal visco-elasto-capillary (TVEC).
The phase before the EC regime ($t - t_{\mathrm{\hspace{1pt} EC}} < 0$) is known as the IC/VC regime, and Newtonian fluids exhibit only the IC/VC regime. 
In contrast, viscoelastic fluids exhibit an EC regime. 
By setting the onset time of the TVEC regime to $t_{\mathrm{FE}}$, the EC regime is defined by the interval $0 \leq t - t_\mathrm{\hspace{1pt} EC} < t_\mathrm{\hspace{1pt} FE} - t_\mathrm{\hspace{1pt} EC}$. 
In the duration extending beyond the EC regime up to the point just before the filament breaks ($t - t_{\mathrm{\hspace{1pt} EC}} \geq t_{\mathrm{\hspace{1pt} FE}} - t_{\mathrm{\hspace{1pt} EC}}$), the TVEC regime can be identified by the rapid reduction in the radius ratio of the filament over time.
The EC regime is theoretically a perfect uniaxial extension with exponential decay of the radius ratio $R/R_{0}$ with time $t - t_{\mathrm{\hspace{1pt} EC}}$, expressed as follows \cite{dinic2019macromolecular,dinic2015extensional,tamano2017dynamics,sur2018drop,mckinley2005visco,wagner2015analytic,prabhakar2006effect,omidvar2019detecting}:
\begin{equation}
\frac{R(t)}{R_{0}} = \left(\frac{GR_{0}}{2\Gamma}\right)^{1/3}\exp\!\left(-\frac{t - t_{\mathrm{\hspace{1pt} EC}}}{3\lambda_\mathrm{E}}\right)
\label{radius in ECregime},
\end{equation}
where $G$, $\Gamma$, and $\lambda_\mathrm{E}$ denote the elastic modulus, surface tension, and extensional relaxation time of the fluid, respectively. 
Moreover, within the EC regime, the extensional stress $\sigma_{\rm \hspace{1pt} E}$ can be expressed as follows:
\begin{equation}
\sigma_{\rm \hspace{1pt} E} = \frac{\Gamma}{R(t)}
\label{stress in ECregime}.
\end{equation}
In uniaxial extensional flow, the extensional stress is almost constant within the EC regime.

\subsection{\label{rheo-optical technique}Rheo-optical technique with high-speed polarization camera}
We employed birefringence measurements to determine the phase retardation of polarized light passing through worm-like micelles under stress loading.
Fig.~\ref{rheo_optical_technique}(b) shows the changes in the polarization state of light when transmitted through a birefringent object.
The incident electromagnetic wave, represented by the red curve, has two orthogonal components, represented by the blue and green curves.
When there is a phase retardation of $\pi/2$ rad between the two components, the incident light is circularly polarized.
When light is transmitted through a birefringent object (worm-like micelles) under stress loading, the propagation speed of each component wave differs because the refractive indices $n_{\parallel}$ and $n_{\perp}$ differ.
Thus, the phases of the components differ, and the retardation values change.

The retardation $\delta$ is given by the integral of the birefringence $\delta n$ over infinitesimal volume elements within the object along the optical axis: $z$ direction (see Eq. (\ref{eq2})).
The high-speed polarization camera captures photoelastic phenomena using a pixelated polarizer array at a frame rate of up to $1.5 \times 10^6$ frames/s.
The array consisted of four adjacent polarizers ($2\times2$) arranged in four orientations.
The instantaneous retardation ${\delta}$ and orientation angle $\varphi$ are calculated from the light intensities captured by the four polarizers ($I_1$ at 0$^\circ$, $I_2$ at 45$^\circ$, $I_3$ at 90$^\circ$, and $I_4$ at 135$^\circ$)\cite{onuma2014development}.
\begin{align}
{\delta} &= \frac{\lambda}{2 \pi} \sin ^{ -1 }{ \frac { \sqrt { { ( { I }_{ 3 }-{ I }_{ 1 } ) }^{ 2 }+{ ( { I }_{ 2 }{-I }_{ 4 } ) }^{ 2 } } }{ ({ I }_{ 1 }+{ I }_{ 2 }+{ I }_{ 3 }+{ I }_{ 4 } )/2} }
\label{delta},
\\[6pt]
\varphi &=\frac{1}{2}\tan ^{ -1 }{ \frac { { I }_{ 3 }-{ I }_{ 1 } }{{ I }_{ 2 }-{ I }_{ 4 }}}
\label{phi},
\end{align}
where $\lambda$ is the wavelength of the incident light.
In the present study, the incident light emitted from a green LED light with a wavelength of 525 nm was polarized by a circular polarization sheet consisting of a polarizer and quarter-wave plate. 
The incident light was directed straight into the camera.
The orientation angle $\varphi$ is defined as the tilt angle of the worm-like micelles with respect to the horizontal ($x$) direction; the vertical ($y$) direction is the extension direction.
Details regarding the spatio-temporal resolution and other specifications of the high-speed polarization camera setup for the experiment are provided in the Appendix.

\subsection{\label{worm-like micellar soludions}Worm-like micellar solutions}
In the present study, eight worm-like micellar solutions with different molar concentrations of CTAB (FUJIFILM Wako Pure Chemical Co.) and NaSal (Sigma-Aldrich Co.) were used, as shown in Table \ref{Table:solutions}.
\begin{table}[b!]
    \centering
    \caption{CTAB/NaSal aqueous solutions with varied concentrations of CTAB ($c_\mathrm{D}$) and NaSal ($c_\mathrm{S}$)  used in the present study.}
    \label{Table:solutions}
    \tabcolsep = 0.2cm
    \renewcommand \arraystretch{1.35}
    \begin{tabular}{c c c c c}
    \hline \hline
        \multirow{2}{*}{Solution No.} & $c_\mathrm{D}$ & $c_\mathrm{S}$ & $c_\mathrm{S}/c_\mathrm{D}$ & $c_\mathrm{S}-c_\mathrm{D}$ \\ 
         & [M] & [M] & [-] & ($\sim c^*_\mathrm{S}$) [M] \\ \hline 
        $\mathrm I$ &  0.0050  & 0.0385 & 7.7 & 0.0335 \\ 
        $\mathrm {I\hspace{-0.6pt}I}$ &  0.0100  & 0.0770 & 7.7 & 0.0670 \\ 
        $\mathrm {I\hspace{-0.6pt}I\hspace{-0.6pt}I}$ &  0.0200  & 0.1540 & 7.7 & 0.1340 \\
        $\mathrm {I\hspace{-0.6pt}V}$ &  0.0300  & 0.2300 & 7.7 & 0.2 \\
        $\mathrm V$ &  0.0075  & 0.2075 & 27.7 & 0.2 \\ 
        $\mathrm {V\hspace{-0.6pt}I}$ &  0.0100  & 0.2100 & 21.0 & 0.2 \\ 
        $\mathrm {V\hspace{-0.6pt}I\hspace{-0.6pt}I}$ &  0.0125  & 0.2125 & 17.0 & 0.2 \\ 
        $\mathrm {V\hspace{-0.6pt}I\hspace{-0.6pt}I \hspace{-0.6pt} I}$
 &  0.0150  & 0.2150 & 14.3 & 0.2 \\ \hline \hline
    \end{tabular}
\end{table}
The molar concentrations of CTAB and NaSal are denoted as $c_\mathrm{D}$ [M] and $c_\mathrm{S}$ [M], respectively. 
Previous studies on rheo-optical measurements of worm-like micellar solutions under shear flow \cite{shikata1994rheo,wheeler1996structure,C_Takahashi,ito2015temporal} employed conditions with $c_\mathrm{D}=0.03$ M and $c_\mathrm{S}=0.23$ M. 
Our study adopted similar conditions as the standard (solution $\mathrm {I\hspace{-0.6pt}V}$), prepared three solutions (solutions $\mathrm I$ to $\mathrm{I\hspace{-0.6pt}I\hspace{-0.6pt}I}$) with a fixed molar ratio of $c_\mathrm{S}/c_\mathrm{D} = 7.7$, and four solutions (solutions $\mathrm V$ to $\mathrm {V\hspace{-0.6pt}I\hspace{-0.6pt}I \hspace{-0.6pt} I}$) with a fixed molar difference of $c_\mathrm{S}-c_\mathrm{D}=0.2$ M, totaling eight solutions. The $c_\mathrm{D}$ values of all solutions exceeded the critical micelle concentration of $9 \times 10^{-4}$ M, at which the surface tension is known to be $\Gamma = 36$ mN/m\cite{cooper2002drop}. 
Using a surface tensiometer (DY-300, Kyowa Interface Science Co., Ltd.) employing the Wilhelmy plate method\cite{gaonkar1987uncertainty}, the surface tensions of all the solutions used in the present study were measured and confirmed to be at $\Gamma = 36$ mN/m. 

\section{\label{result}Results and discussion \protect}
This section describes obtaining the extensional stress and birefringence through rheo-optical measurements of worm-like micellar solutions uniaxially extended by the liquid dripping method. 
In addition, it describes the calculation of the stress-optical coefficient.

\subsection{\label{Extensional stress}Extensional behavior and stress}
\begin{figure*}[t!]
\includegraphics[width=1\textwidth]{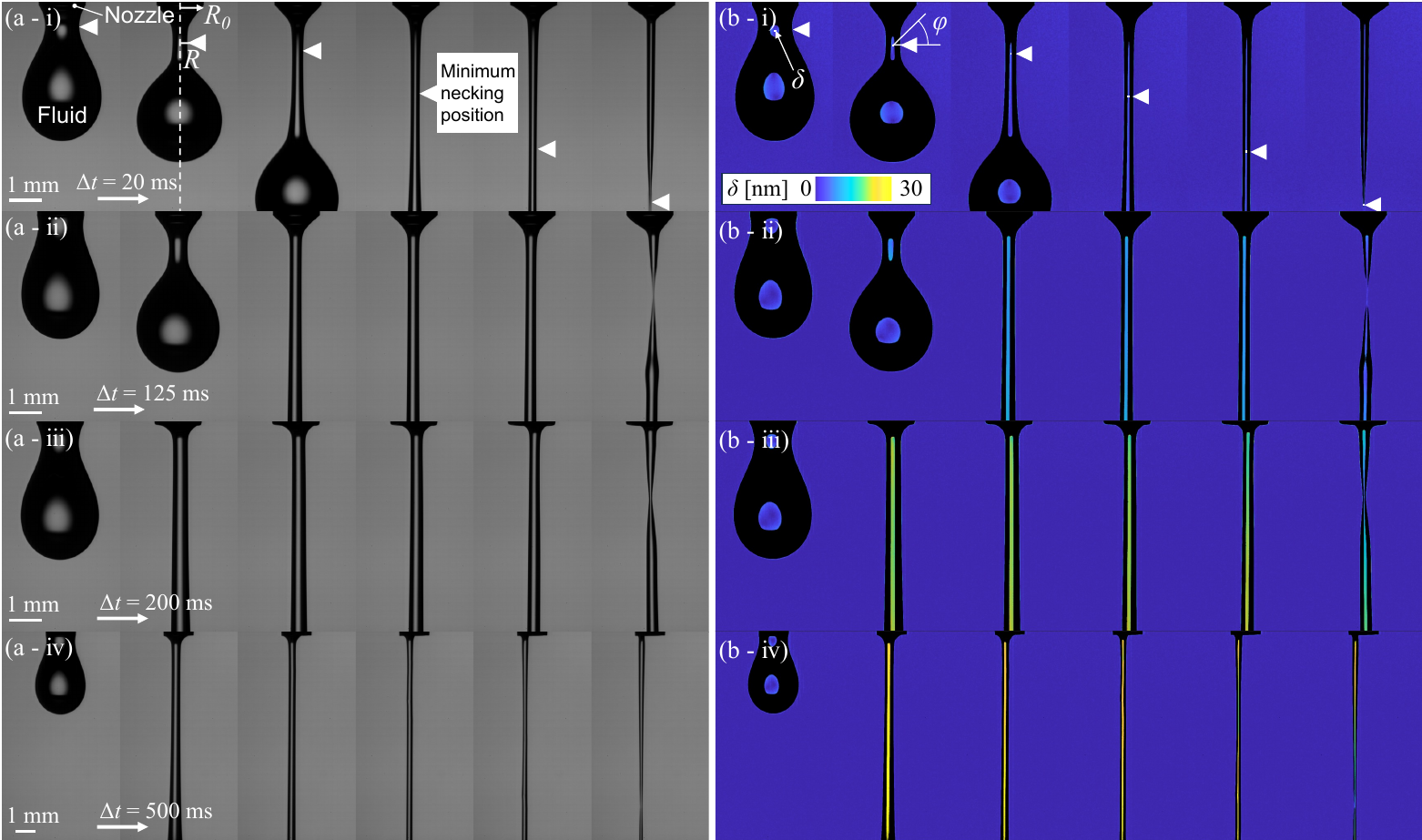}
\caption{Temporal images of light intensity and phase retardation of uniaxially extending worm-like micellar solutions at a constant molar ratio ($c_\mathrm{S} / c_\mathrm{D} = 7.7$). The left column contains (a) light intensity images, and the right column contains (b) phase retardation images, corresponding to different solutions: (i) \ solution $\mathrm I$, (ii) \ solution $\mathrm {I\hspace{-0.6pt}I}$, (iii) \ solution $\mathrm {I\hspace{-0.6pt}I\hspace{-0.6pt}I}$, (iv) \ solution $\mathrm {I\hspace{-0.6pt}V}$. $\Delta \hspace{0.4pt} t$ indicates time interval between each image. The original images captured by the high-speed polarized camera had a resolution of 1024×1024 pixels. The images were cropped to 575×1024 pixels to exclude areas not showing the droplets. Furthermore, the deep blue background in the phase retardation images does not contribute to the contours because the orientation angle was random and the measured phase retardation was zero.}
\label{intensity_retardation_images}
\end{figure*}

Temporal images of the light intensity and phase retardation of CTAB/NaSal solutions $\mathrm I$ to $\mathrm {I\hspace{-0.6pt}V}$ at various concentrations while maintaining a constant molar ratio ($c_\mathrm{S}/c_\mathrm{D}=7.7$) are shown in Fig. \ref{intensity_retardation_images}.
The minimum radius $R$ of the filaments must be measured to investigate the extensional behavior of the filaments quantitatively, as indicated by the white arrow in Fig. \ref{intensity_retardation_images}(a-i). 
The temporal development of the extended liquid filament shows the thinning characteristics of viscoelastic fluids under all concentrations.
The time taken for these extended filaments to break increases with increasing molar concentrations. 
These trends are also confirmed in solutions $\mathrm V$ to $\mathrm {V\hspace{-0.6pt}I\hspace{-0.6pt}I \hspace{-0.6pt} I}$, where the molar difference is kept constant ($c_\mathrm{S}-c_\mathrm{D}=0.2$ M), as detailed in the Appendix.
Further discussion of the phase retardation images (Fig. \ref{intensity_retardation_images}(b)) is provided in Section III.B.

\begin{figure}[t!]
\includegraphics[width=\columnwidth]{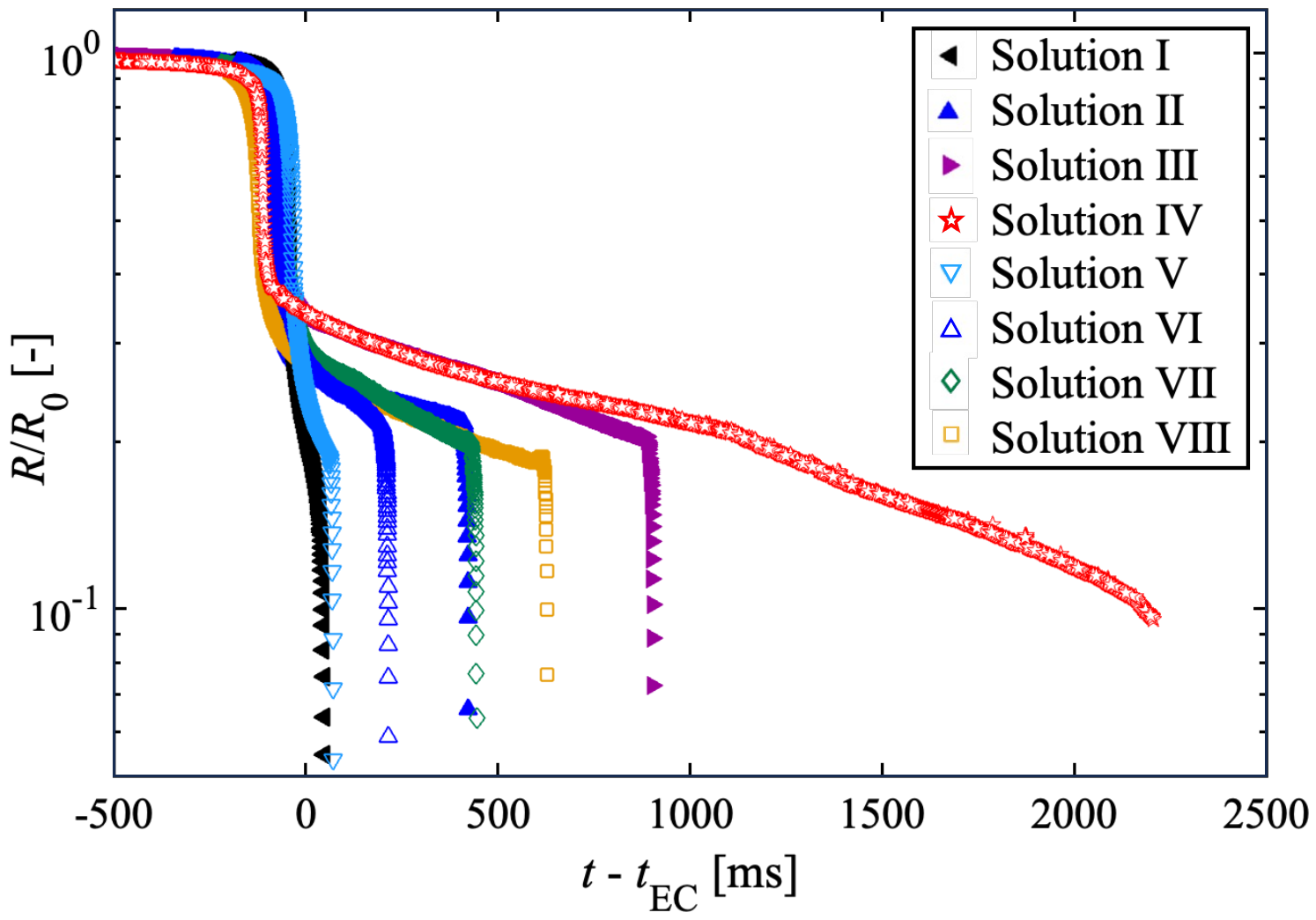}
\caption{Temporal evolution of radius ratio $R/R_0$ for worm-like micellar solutions $\mathrm I - \mathrm {V\hspace{-0.6pt}I\hspace{-0.6pt}I \hspace{-0.6pt} I}$ (see Table \ref{Table:solutions}).}
\label{radius_ratio}
\end{figure}

A semi-logarithmic graph of the filament radius ratio $R/R_0$, obtained by analyzing the light intensity images (Fig. \ref{intensity_retardation_images}(a)), is shown in Fig. \ref{radius_ratio}. 
For all concentrations, after time $t - t_{\mathrm{EC}} = 0$ ms, the radius ratio transitioned into the EC regime, exhibiting exponential decay over time. 
Moreover, IC/VC, EC, and TVEC regimes are confirmed for all solutions except Solution IV. 
However, the TVEC regime is not observed for solution $\mathrm {I\hspace{-0.6pt}V}$ because their magnification level during the experiments was lower than that of the other solutions. 
This resulted in reduced spatial resolution (see Table \ref{Table:camera} in the Appendix), which was inadequate to capture the breakage of the extending filaments.

Table \ref{Table:C} lists the extension time $t_{\mathrm{FE}}-t_{\mathrm{EC}}$ [ms] for each solution, representing the EC regime duration (Fig. \ref{radius_ratio}). 
\begin{table}[b!]
    \centering
    \caption{Extensional time range $t_\mathrm{\hspace{1pt} FE}-t_\mathrm{\hspace{1pt} EC}$, phase retardation at onset time of EC regime $\delta_\mathrm{\hspace{1pt} EC}$, and stress-optical coefficient $C$ under extensional flow.}
    \label{Table:C}
    \tabcolsep = 0.2cm
    \renewcommand \arraystretch{1.35}
    \begin{tabular}{c c c c}
    \hline \hline
        Solution No. & $t_\mathrm{\hspace{1pt} FE}-t_\mathrm{\hspace{1pt} EC}$ [ms] & $\delta_\mathrm{\hspace{1pt} EC}$ [nm] & $C$ $\times 10^{7}$ [Pa$^{-1}$] \\ \hline
        $\mathrm I$ & 60 & 3.9 & $-0.58 \pm 0.03$ \\ 
        $\mathrm {I\hspace{-0.6pt}I}$ & 410 & 10.9 & $-1.41 \pm 0.05$ \\ 
        $\mathrm {I\hspace{-0.6pt}I\hspace{-0.6pt}I}$ & 880 & 23.0 & $-2.71 \pm 0.21$ \\ 
        $\mathrm {I\hspace{-0.6pt}V}$ & 1850 & 29.6 & $-3.51 \pm 0.44$ \\ 
        $\mathrm V$ & 65 & 6.5 & $-0.90 \pm 0.05$ \\ 
        $\mathrm {V\hspace{-0.6pt}I}$ & 200 & 10.9 & $-1.44 \pm 0.06$ \\ 
        $\mathrm {V\hspace{-0.6pt}I\hspace{-0.6pt}I}$ & 430 & 14.3 & $-1.78 \pm 0.14$ \\ 
        $\mathrm {V\hspace{-0.6pt}I\hspace{-0.6pt}I \hspace{-0.6pt} I}$
 & 600 & 15.8 & $-1.93 \pm 0.11$ \\ \hline \hline
    \end{tabular}
\end{table}
The table also includes the phase retardation $\delta_\mathrm{\hspace{1pt} EC}$ [nm] at the onset of the EC regime and stress-optical coefficient $C$ [Pa$^{-1}$], discussed in Section III.B. 
The extension time increases with CTAB concentration $c_\mathrm{D}$ owing to the increase in the total worm-like micelles in the solution. 
Shikata et al.\cite{shikata1987micelle,shikata1988micelle} experimentally demonstrated that worm-like micelles in CTAB/NaSal solutions were formed at a 1:1 molar ratio of CTAB to NaSal. 
Therefore, in solutions with $c_\mathrm{S}-c_\mathrm{D}>0$, the $c_\mathrm{D}$ value indicates the total quantity of worm-like micelles. 
In the present study, an increase in $c_\mathrm{D}$, corresponding to the total worm-like micelles, increased the complex entanglements among the worm-like micelles under extensional stress loading, prolonging extension time ($t_{\mathrm{FE}}-t_{\mathrm{EC}}$).

Furthermore, comparing the extension times in Table \ref{Table:C} for solutions $\mathrm {I\hspace{-0.6pt}I}$ (constant molar ratio) and $\mathrm {V\hspace{-0.6pt}I}$ (constant molar difference), both with $c_\mathrm{D} = 0.0100$ M, reveals that solution $\mathrm {V\hspace{-0.6pt}I}$ has a shorter extension time. 
This is due to the differences in the molar concentrations of free salicylate anions ($c^*_\mathrm{S}$ [M]) between the two solutions. 
Free salicylate anions serve as catalysts for the splitting reaction of worm-like micelles in CTAB/NaSal solutions. 
Their concentrations ($c^*_\mathrm{S}$) are nearly equivalent to the difference between the concentrations of CTAB and NaSal, represented as $c_\mathrm{S}-c_\mathrm{D}$ \cite{shikata1988micelle}.
Assuming $c^*_\mathrm{S} \sim c_\mathrm{S} - c_\mathrm{D}$, the free salicylate anion concentration was found to be approximately $c^*_\mathrm{S} \sim 0.0670$ and $c^*_\mathrm{S} \sim 0.2000$ M for solutions $\mathrm {I\hspace{-0.6pt}I}$ and $\mathrm {V\hspace{-0.6pt}I}$, respectively. An increase in the free salicylate anion concentration affected the formation capability and stability of micelles. Therefore, solution $\mathrm {V\hspace{-0.6pt}I}$, with a higher free salicylate anion concentration, likely exhibited reduced viscoelasticity compared to solution $\mathrm {I\hspace{-0.6pt}I}$, reducing the extension time.

The extensional stress $\sigma_{\rm \hspace{1pt} E}$ of each solution is calculated by substituting the measured minimum radius $R$ of the filament into Eq. (\ref{stress in ECregime}). 
\begin{figure*}[t!]
\includegraphics[width=\textwidth]{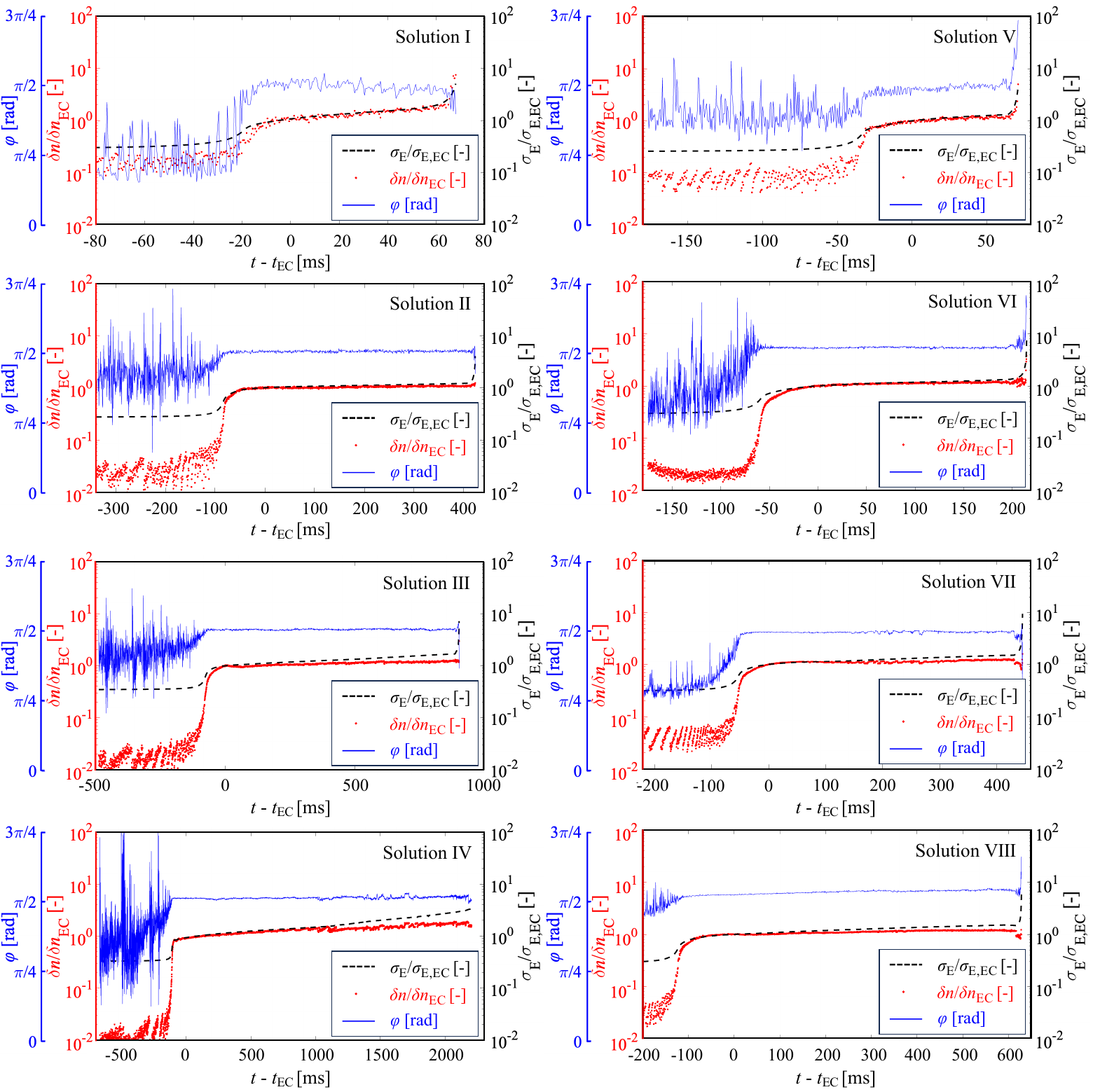}
\caption{Temporal evolution of normalized extensional stress $\sigma_{\rm \hspace{1pt} E}/\sigma_{\rm \hspace{1pt} E,EC}$, normalized birefringence $\delta n/\delta n_\mathrm{\hspace{1pt} EC}$, and orientation angle $\varphi$ for solutions $\mathrm I - \mathrm {V\hspace{-0.6pt}I\hspace{-0.6pt}I \hspace{-0.6pt} I}$.}
\label{Normalized_data}
\end{figure*}
Figure \ref{Normalized_data} shows the normalized extensional stress $\sigma_{\rm \hspace{1pt} E}/\sigma_{\rm \hspace{1pt} E,EC}$ for all solutions. 
Here, $\sigma_{\rm \hspace{1pt} E,EC}$ represents the extensional stress at the onset of the EC regime ($t - t_\mathrm{\hspace{1pt} EC} = 0$). 
Additionally, the figure includes the normalized birefringence $\delta n/\delta n_\mathrm{\hspace{1pt} EC}$ and orientation angle $\varphi$, discussed in Section III.B.
For all solutions, the normalized extensional stress increases in the IC/VC regime ($t - t_\mathrm{\hspace{1pt} EC} < 0$), remains roughly constant within the EC regime ($0 \leq t - t_\mathrm{\hspace{1pt} EC} < t_\mathrm{\hspace{1pt} FE} - t_\mathrm{\hspace{1pt} EC}$), and then sharply increases in the TVEC regime ($t - t_\mathrm{\hspace{1pt} EC} \geq t_\mathrm{\hspace{1pt} FE} - t_\mathrm{\hspace{1pt} EC}$). 
Theoretically, within the EC regime, where the elastic force balances the capillary force, the filament of a viscoelastic fluid extends at a constant strain rate, generating a constant extensional stress acting on it \cite{mckinley2005visco}. 
Therefore, the present measurements of extensional stress  agree with this theory, confirming its validity.

\subsection{\label{birefringence result}Birefringence and orientation angle}
The measured data for birefringence and orientation angles (Fig. \ref{Normalized_data}) were extracted by analyzing the phase retardation images (Fig. \ref{intensity_retardation_images}(b)). 
The phase retardation and orientation angle obtained using the proposed method represent the spatial average values at the central part of the filament at the minimum radius position, as indicated by the white square in Fig. \ref{intensity_retardation_images}(b-i). 
Analyzing the central part is crucial to avoid the effects of refraction and scattering of transmitted light owing to the filament's curvature, as detailed in the Appendix. 
The measured birefringence data were obtained by dividing the measured phase retardation data by the filament diameter ($\delta n = \delta/2R$).
In addition to the normalized extensional stress, Fig. \ref{Normalized_data} also shows the temporal evolution of normalized birefringence and orientation angle. 
The following discussion explains these results.

Orientation angle $\varphi$ (blue line in Fig. \ref{Normalized_data}) is based on a horizontal reference of $0$ rad, with the worm-like micelles oriented in the direction of extension at $\pi/2$ rad.
At the initial stage of extension in the IC/VC regime ($t - t_\mathrm{\hspace{1pt} EC} < 0$), $\varphi$  oscillated numerically and was unstable across all concentrations, indicating a statistically random orientation of the worm-like micelles.
However, immediately before entering the EC regime, $\varphi$ begins to converge, whereas within the EC regime ($0 \leq t - t_\mathrm{\hspace{1pt} EC} < t_\mathrm{\hspace{1pt} FE} - t_\mathrm{\hspace{1pt} EC}$) it settles at a constant $\pi/2$ rad. 
Therefore, within the EC regime, where the orientation angle remains constant at $\pi/2$ rad, the worm-like micelles are statistically highly oriented in the direction of extension, validating the applicability of the stress-optical rule (Eq. (\ref{eq4})).
Interestingly, the convergence of the orientation angle towards a constant value ($\pi/2$ rad) began even before the EC regime, indicating that the orientation of the worm-like micelles towards the extension direction began earlier than the uniaxially extended state guaranteed in the EC regime. 
In the TVEC regime ($t - t_\mathrm{\hspace{1pt} EC} \geq t_\mathrm{\hspace{1pt} FE} - t_\mathrm{\hspace{1pt} EC}$), where the filament extends further until breaking, the orientation angle becomes unstable because the effect of light refraction due to the curvature of the liquid filament becomes more pronounced.
Note that not all worm-like micelles in a solution are oriented uniformly, rather they are statistically highly oriented.

The orientation direction of the worm-like micelles being the same as the filament extension direction led to the following two important findings.
First, the measured birefringence is obtained as the intrinsic birefringence.
The relation between the flow-induced birefringence $\delta n$ and orientation angle $\varphi$ is expressed as follows \cite{okada2016reliability}:
\begin{equation}
\delta n = f \delta n^{0} = \frac{3\cos^{2}(\pi/2-\varphi)-1}{2} \delta n^{0}
\label{orientation_angle},
\end{equation}
where $\delta n^{0}$ is the intrinsic birefringence, which is a material constant for the birefringence of fully oriented worm-like micelles.
Importantly, intrinsic birefringence ($f \sim 1$) is considered within the EC regime because the orientation angle remains constant at $\pi/2$.
Second, 
because the direction of the extensional stress ($\sigma_{\perp}$) aligns with the refractive index ($n_{\perp}$), it follows from Eq. (\ref{eq1}) that $\delta n = -n_{\perp} < 0$, indicating negative birefringence. Therefore, the stress-optical coefficient ($C$) for the CTAB/NaSal solutions is always negative, which is consistent with previous studies \cite{shikata1994rheo,wheeler1996structure,C_Takahashi}.

To validate the stress-optical rule (Eq. (\ref{eq4})), the match between the normalized extensional stress and birefringence ($\sigma_{\rm \hspace{1pt} E}/\sigma_{\rm \hspace{1pt} E,EC} = \delta n/\delta n_\mathrm{\hspace{1pt} EC}$) must be verified. 
By normalizing the measured birefringence with the birefringence at the onset of the EC regime, the normalized birefringence $\delta n/\delta n_\mathrm{\hspace{1pt} EC}$ was calculated and compared with the normalized extensional stress $\sigma_{\rm \hspace{1pt} E}/\sigma_{\rm \hspace{1pt} E,EC}$.
In Fig. \ref{Normalized_data}, across all concentrations, normalized extensional stress and birefringence exhibit an increase just before entering the EC regime and tend to converge to the corresponding constant values within the EC regime. 
Therefore, in the EC regime, the stress-optical rule (Eq. (\ref{eq4})) is sufficiently satisfied, allowing the calculation of the stress-optical coefficient $C$.

In Fig. \ref{Normalized_data}, the orientation angle for solution $\mathrm I$ ($c_\mathrm{D}=0.0050$ M, $c_\mathrm{S}=0.0385$ M) does not fully converge to $\pi/2$ rad and shows significant fluctuations within the EC regime.
This is because solution $\mathrm I$, having a lower CTAB concentration ($c_\mathrm{D}$) than the other solutions, results in a smaller total quantity of worm-like micelles, making the measurement challenging.
The results confirm that solution $\mathrm I$ exhibits a small phase retardation of $\delta _\mathrm{\hspace{1pt} EC}=3.9$ nm (Table \ref{Table:C}). 
Considering that the measurement limit for phase retardation using the high-speed polarization camera is approximately 5 nm, this camera cannot accurately capture the minor phase retardation induced by birefringence in solution $\mathrm I$.

\subsection{\label{birefringence}Stress-optical coefficient}
This subsection discusses the stress-optical coefficient, calculated using the measured extensional stress and birefringence. 
To demonstrate the validity of the stress-optical rule under shear flow, Takahashi et al. \cite{C_Takahashi} reported the need to verify two distinct phenomena: (a) proportionality between the refractive index and stress tensors and (b) constancy of the stress-optical coefficient regardless of the shear rate. 
Phenomena (a) was verified by demonstrating the alignment of the orientation direction of worm-like micelles with the direction of extensional stress, as discussed in Section III.B.
However, to confirm the validity of the stress-optical coefficients obtained through the present method, the validity of (b), even under extensional flow, must be investigated.
Note that the present study employs the "extension rate" instead of the "shear rate" used in (b).
The verification results are discussed below.

\begin{figure}[tb!]
\includegraphics[width=\columnwidth]{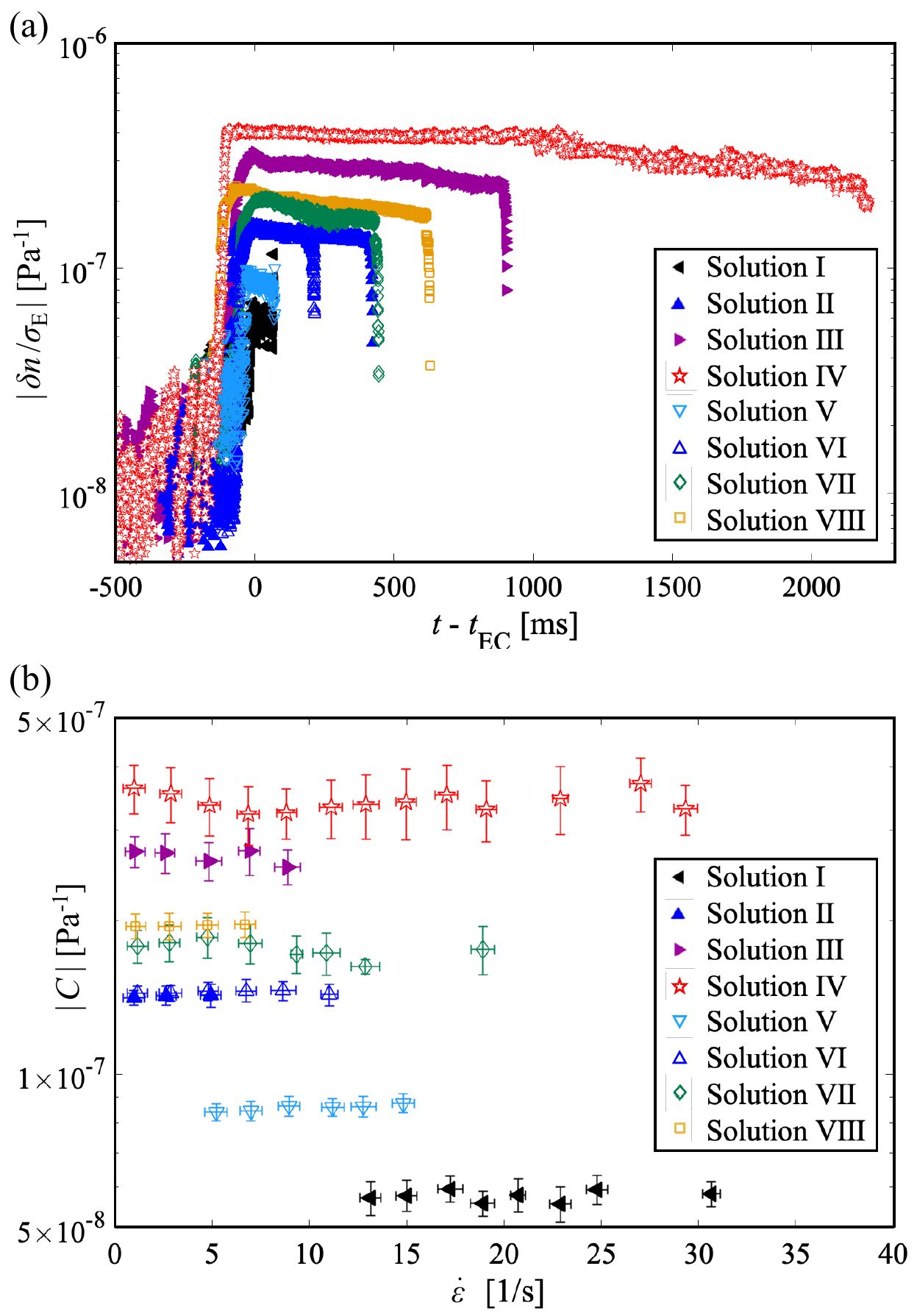}
\caption{Graphs of the ratio of birefringence and extensional stress $\delta n / \sigma_{\rm \hspace{1pt} E}$, and stress-optical coefficient $C$ of worm-like micellar solutions under uniaxial extensional stress condition. (a) $|\delta n / \sigma_{\rm \hspace{1pt} E}|$ vs. whole ranges of time $t - t_\mathrm{\hspace{1pt} EC}$. (b) $|C|$ vs. extensional rate $\dot{\varepsilon}$ within EC regime.}
\label{C}
\end{figure}

Figure \ref{C}(a) shows the ratio of birefringence to extensional stress $|\delta n / \sigma_{\rm \hspace{1pt} E}|$ [Pa$^{-1}$] for solutions $\mathrm I$ to $\mathrm {V\hspace{-0.6pt}I\hspace{-0.6pt}I \hspace{-0.6pt} I}$. 
The horizontal axis represents time $t-t_\mathrm{\hspace{1pt} EC}$ [ms]. 
During the initial stages of extension in the IC/VC regime ($t - t_\mathrm{\hspace{1pt} EC} < 0$) and just before the filament breakage in the TVEC regime ($t - t_\mathrm{\hspace{1pt} EC} \geq t_\mathrm{\hspace{1pt} FE} - t_\mathrm{\hspace{1pt} EC}$), the ratio $|\delta n / \sigma_{\rm \hspace{1pt} E}|$ shows unstable behavior and does not converge across different concentrations.
However, within the EC regime ($0 \leq t - t_\mathrm{\hspace{1pt} EC} < t_\mathrm{\hspace{1pt} FE} - t_\mathrm{\hspace{1pt} EC}$), the ratio $|\delta n / \sigma_{\rm \hspace{1pt} E}|$ stabilizes, confirming the proportional relation between birefringence and extensional stress. 
This observation agrees with the finding that the stress-optical rule is valid only within the EC regime (as mentioned in Section III.B), indicating that the stress-optical coefficient can be considered an intrinsic material property determined within the EC regime.

Figure \ref{C}(b) shows the stress-optical coefficient $|C|$ [Pa$^{-1}$] derived from $|\delta n / \sigma_{\rm \hspace{1pt} E}|$ within the EC regime ($0 \leq t - t_\mathrm{\hspace{1pt} EC} < t_\mathrm{\hspace{1pt} FE} - t_\mathrm{\hspace{1pt} EC}$) for each solution.
The horizontal axis represents the extension rate $\dot{\varepsilon}$ [s$^{-1}$]. 
The plots and error bars indicate the mean $|C|$ and standard deviation of each plot within a bandwidth of 2 s$^{-1}$ for the extension rate.
The constancy of the stress-optical coefficient $|C|$ regardless of the extension rate $\dot{\varepsilon}$ indicates that the stress-optical coefficient can be accurately determined as the intrinsic material property. 
Similar to the stress-optical coefficient under a shear flow being independent of the shear rate and constant, the stress-optical coefficient under uniaxial extensional flow is independent of the extension rate.
The mean and standard deviations of the stress-optical coefficient are listed in Table \ref{Table:C}. The small standard deviations suggest the high reliability of the present method.

The measured stress-optical coefficients under uniaxial extensional flow for solution $\mathrm {I\hspace{-0.6pt}V}$ ($c_\mathrm{D}=0.0300$ M and $c_\mathrm{S}=0.2300$ M) were compared and validated against that under shear flow considering similar concentrations of worm-like micellar solutions from previous studies \cite{shikata1994rheo,wheeler1996structure,C_Takahashi}. 
In the present study, the measured stress-optical coefficient under uniaxial extensional flow is $C=-3.51 \times 10^{-7}$ Pa$^{-1}$ (Table \ref{Table:C}). 
This value is closest to that reported by Wheeler et al. \cite{wheeler1996structure} ($C=-3.49 \times 10^{-7}$ Pa$^{-1}$) and approximately agrees with the values reported by Shikata et al. \cite{shikata1994rheo} and Takahashi et al. \cite{C_Takahashi} under shear flow ($C=-3.1 \times 10^{-7}$ Pa$^{-1}$ and $C=-2.77 \times 10^{-7}$ Pa$^{-1}$, respectively). 
Therefore, the stress-optical coefficient of worm-like micellar solutions is consistent under shear and extensional flows.
Takahashi et al. \cite{C_Takahashi} noted that the stress-optical rule breaks down in shear flow measurements at shear rates below $\dot{\gamma} >$ 4 s$^{-1}$. 
In contrast, our results indicate that the stress-optical rule does not break down. 
Moreover, the stress-optical coefficient measurements for solution $\mathrm {I\hspace{-0.6pt}V}$ are feasible within an extension rate range of $0<\dot{\varepsilon}<30$ s$^{-1}$.

\begin{figure}[b!]
\includegraphics[width=\columnwidth]{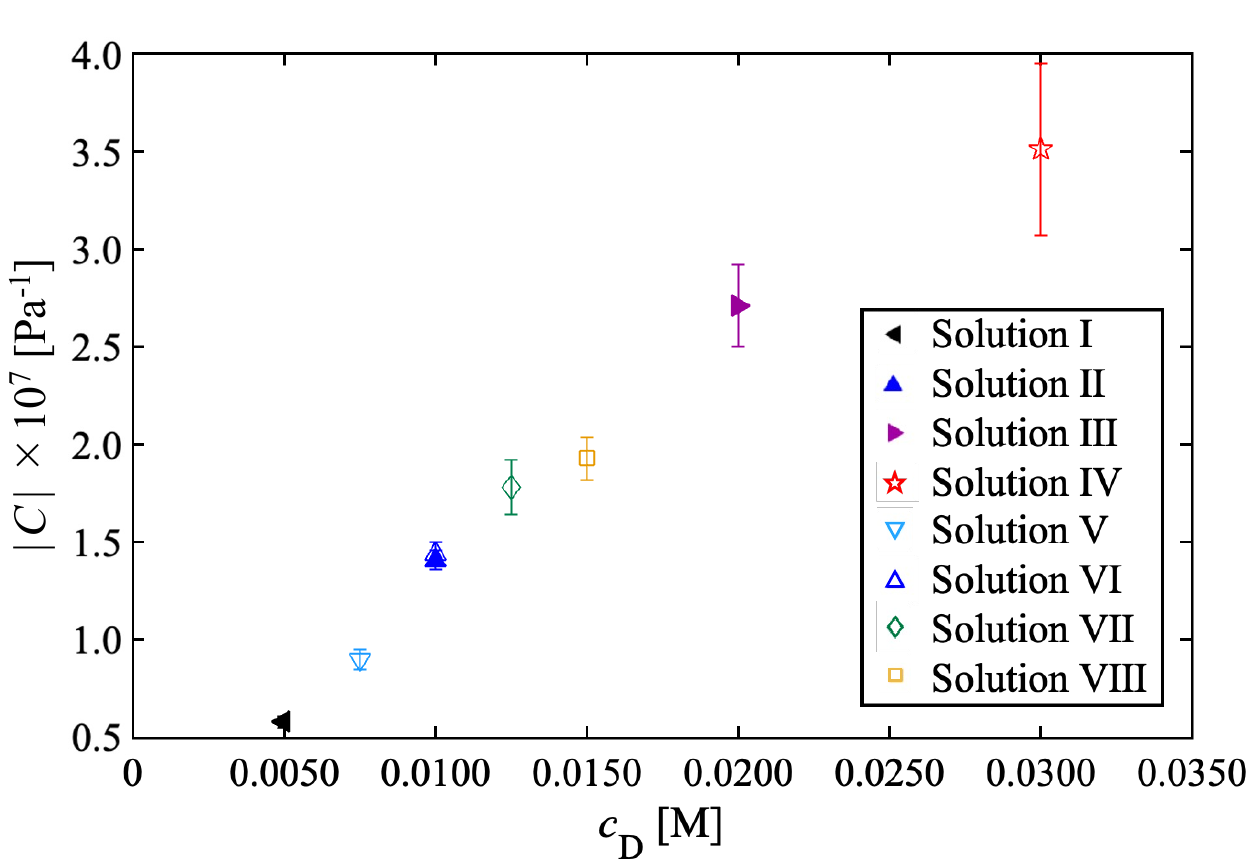}
\caption{Dependence of stress-optical coefficient $|C|$ on CTAB concentration $c_\mathrm{D}$.}
\label{cD_C}
\end{figure}

The concentration dependence of the stress-optical coefficient is also discussed. 
In Fig. \ref{cD_C}, as the CTAB concentration increases within $0.0050 \leq c_\mathrm{D} \leq 0.0300$ M, the absolute value of the stress-optical coefficient monotonically increases. 
Therefore, the stress-optical coefficient of uniaxially extending worm-like micellar solutions depends on the CTAB concentration, that is, the total amount of worm-like micelles in the solution.
Furthermore, the stress-optical coefficients of solutions $\mathrm {I\hspace{-0.6pt}I}$ and $\mathrm {V\hspace{-0.6pt}I}$, having identical CTAB concentration of $c_\mathrm{D}=0.0100$ M, were compared. 
Despite the differences in NaSal concentration ($c_\mathrm{S}$) and free salicylate anion concentration ($c^*_\mathrm{S}$), the stress-optical coefficients of both solutions were nearly equal. 
Hence, we infer that the stress-optical coefficient depends on the total number of micelles ($c_\mathrm{D}$) and is independent of the free salicylate anion concentration ($c^*_\mathrm{S}$) under uniaxial extensional flow.
Shikata et al. \cite{shikata1994rheo} had reported that the stress-optical coefficient for worm-like micellar solutions of different CTAB concentrations ($c_\mathrm{D}$) and a constant free salicylate anion concentration ($c^*_\mathrm{S}$), within $0.01<c_\mathrm{D}<0.1$ M and $0.15<c_\mathrm{S}<0.4$ M, does not depend on ($c_\mathrm{D}$) under shear flow.
Therefore, the concentration dependency of the stress-optical coefficient in worm-like micellar solutions may differ between extensional and shear flows.

The above discussions confirm that the obtained stress-optical coefficients of worm-like micellar solutions under uniaxial extensional flow agree with those under shear flow. 
However, the dependence of the stress-optical coefficients on the CTAB/NaSal concentration, especially the CTAB concentration ($c_\mathrm{D}$), differs between the shear and extensional flows.
One reason may be the difference in the stress applied during extension and shear. 
In addition, as mentioned in Section I, regarding the advantages of uniaxial extensional flow measurements, the high alignment of the index ellipsoids along the extension direction may support more accurate measurements of the stress-optical coefficients than those under shear flow.
Therefore, we believe that the difference in measurement accuracy causes the difference between the shear and extensional flows.

\section{\label{conclusion}Conclusion \protect}
To investigate unsteady changes in the microstructure of worm-like micelles under extensional stress loading, we developed a simple rheo-optical system based on the liquid dripping method and birefringence measurements.
The CTAB/NaSal solutions used as worm-like micellar solutions in the present study were uniaxially extended with the liquid dripping method, and the phase retardation field of the liquid filament under extensional stress loading was measured using a high-speed polarization camera.

The novel rheo-optical technique successfully visualized temporally developing images of the birefringence field and orientation angle of uniaxially extending worm-like micellar solutions induced by the orientation of micelles toward the extensional direction.
This technique provides a linear relation between extensional stress and birefringence, and thus, it can be used as a significant tool for investigating the rheological and optical properties of complex fluids under uniaxial extensional stress loading.

Furthermore, the measured extensional stress and birefringence data enables analyzing the stress-optical coefficient under uniaxial extensional flow.
The stress-optical coefficient was found to be consistent under both shear and extensional flows, indicating the robustness of the stress-optical rule in the CTAB/NaSal solutions.
Moreover, the stress-optical coefficient of uniaxially extending CTAB/NaSal solutions depended on the concentration of CTAB, i.e., the total amount of worm-like micelles in the solutions.

\begin{acknowledgments}
This work was supported by the Japan Society for the Promotion of Science, KAKENHI Grant Nos. 23H01343 and 20K14646, and partially based on the results obtained from the project, JPNP20004, subsidized by the New Energy and Industrial Technology Development Organization (NEDO).
\end{acknowledgments}

\section*{Author declarations}

\subsection*{Conflict of Interest}
The authors have no conflict of interest.

\section*{Data availability}
Data supporting the findings of the present study are available from the corresponding authors upon reasonable request.

\appendix*

\section{\label{morusa} Temporal images of light intensity and phase retardation maintaining a constant molar difference}
\begin{figure*}[t!]
\includegraphics[width=1\textwidth]{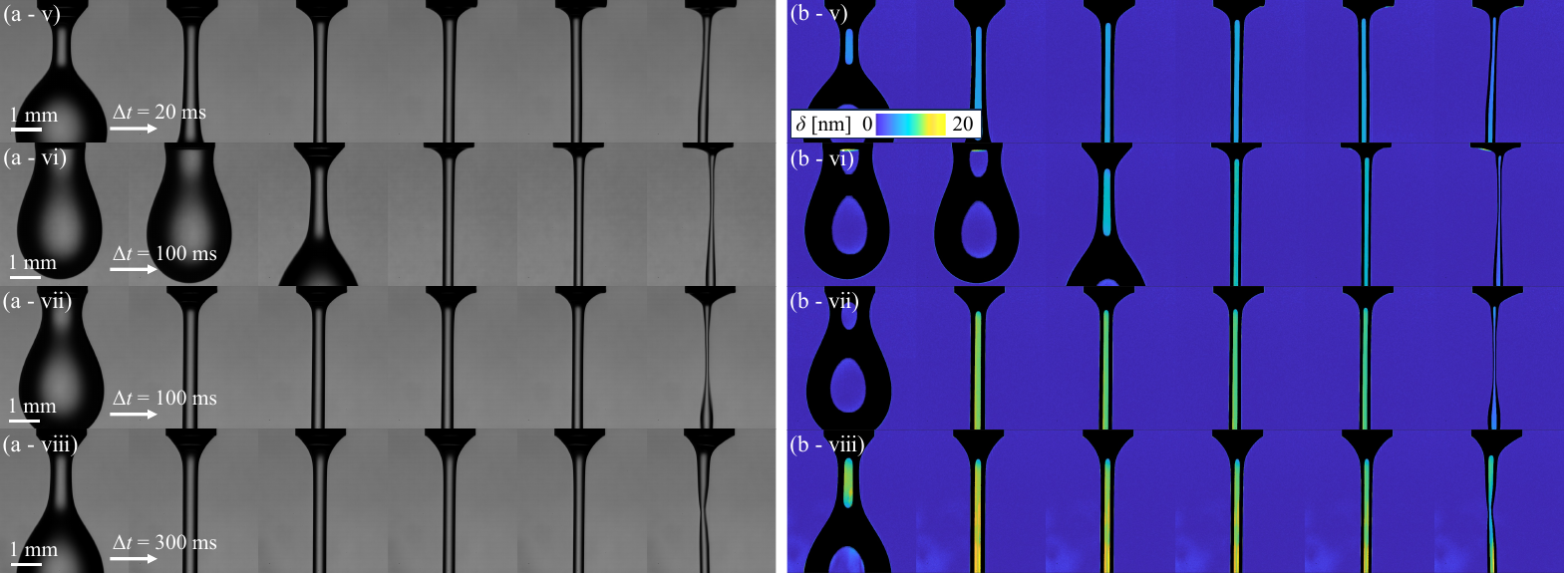}
\caption{Temporal images of light intensity and phase retardation of uniaxially extending worm-like micellar solutions at a constant molar difference ($c_\mathrm{S} - c_\mathrm{D} = 0.2$ M). The left and right column contains (a) light intensity and  (b) phase retardation images, respectively, corresponding to different solutions: (v) \ solution $\mathrm V$, (vi) \ solution $\mathrm {V\hspace{-0.6pt}I}$, (vii) \ solution $\mathrm {V\hspace{-0.6pt}I\hspace{-0.6pt}I}$, (viii) \ solution $\mathrm {V\hspace{-0.6pt}I\hspace{-0.6pt}I\hspace{-0.6pt}I}$. $\Delta \hspace{0.4pt} t$ indicates the time interval between each image.}
\label{intensity_retardation_images_molar_difference}
\end{figure*}

Figure \ref{intensity_retardation_images_molar_difference} shows the light intensity and phase retardation images for solutions $\mathrm V$ to $\mathrm {V\hspace{-0.6pt}I\hspace{-0.6pt}I\hspace{-0.6pt}I}$ having a constant molar difference ($c_\mathrm{S} - c_\mathrm{D} = 0.2$ M).
As the concentration of CTAB/NaSal increases, an increase in the time until the extending filament breaks and phase retardation is confirmed.
These trends are similar to those observed for solutions with a constant molar ratio, as shown in Fig. \ref{intensity_retardation_images}.

\section{\label{area} Shooting conditions}
\begin{figure}[t!]
\includegraphics[width=0.86\columnwidth]{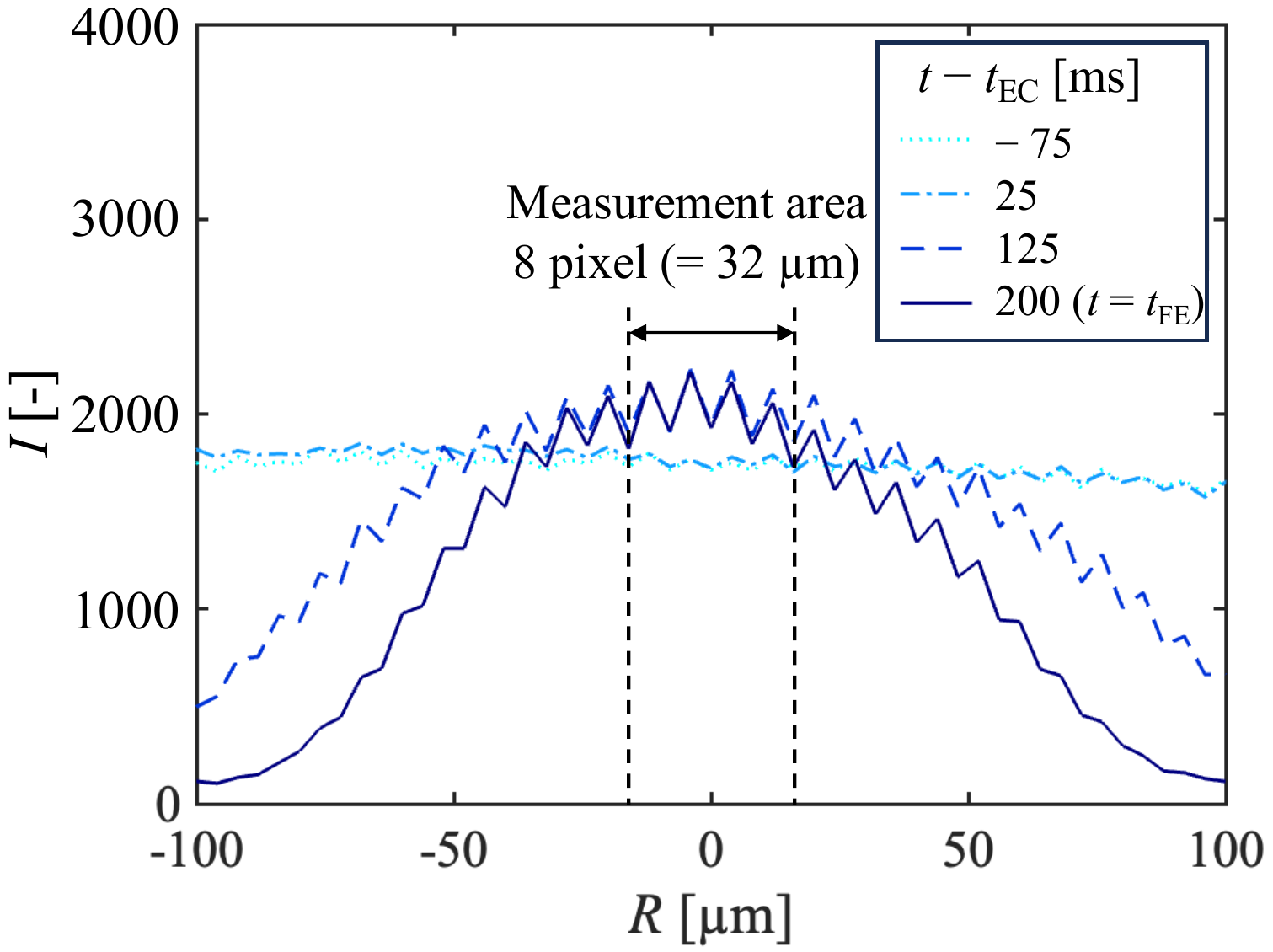}
\caption{Line profile of light intensity distribution at the center of liquid filament for solution $\mathrm {V\hspace{-0.6pt}I}$.}
\label{line_profile}
\end{figure}

Table \ref{Table:camera} lists the shooting conditions of the proposed method. 
\begin{table}[b!]
    \centering
    \caption{Shooting conditions.}
    \label{Table:camera}
    \renewcommand \arraystretch{1.35}
    \begin{tabular}{lccc}
        \hline\hline
         Solution number& $\mathrm I$ to $\mathrm {I\hspace{-0.6pt}I\hspace{-0.6pt}I}$ & $\mathrm {I\hspace{-0.6pt}V}$ & $\mathrm V$ to $\mathrm {V\hspace{-0.6pt}I\hspace{-0.6pt}I \hspace{-0.6pt} I}$\\ \hline
        Shooting area [pixels] & 1024 × 1024  & 1024 × 1024 & 1024 × 1024 \\ 
        Magnification & 3.125 & 2.00 & 5.00\\
        Shooting area [mm] & 6.65 × 6.65  & 10.24 × 10.24 & 4.06 × 4.06 \\ 
        Spatial resolution & \multirow{2}{*}{6.50} & \multirow{2}{*}{10.0} & \multirow{2}{*}{3.97} \\[-1mm]
        [µm/pixel] & & & \\
        Frame rate [fps] & 2000  & 2000 & 2000 \\       
        Shutter speed [s] & 1/2000  & 1/2000 & 1/2000 \\ 
        Analysis area of & \multirow{2}{*}{4 × 4} & \multirow{2}{*}{4 × 4} & \multirow{2}{*}{8 × 8} \\[-1mm]
        $\delta$ and $\varphi$ [pixels] & & & \\ \hline\hline
    \end{tabular}
\end{table}
For determining these conditions, it is necessary to observe the temporal evolution of the filament's extension behavior as well as measure phase retardation $\delta$ and orientation angle $\varphi$ at the filament center with high spatial and temporal resolutions. 
Because the filament has a cylindrical cross-section, scattering occurs when polarized light from the light source passes through the curved parts during side observation, preventing accurate measurement.
Therefore, in the central part of the filament, the areas unaffected by curvature were analyzed to measure the phase retardation and orientation angle. 
This analysis utilized light intensity images to examine the areas where the light intensity line profile at the filament's minimum radius position is constant. 
As an example, Fig. \ref{line_profile} shows the line profile for solution $\mathrm {V\hspace{-0.6pt}I}$.
The light intensity remains constant within a central area of 8 pixels (= 32 µm) at each time point, suggesting that the area is unaffected by the filament's curvature. Therefore, to measure the phase retardation and orientation angle in solution $\mathrm {V\hspace{-0.6pt}I}$, an analysis area of 8 × 8 pixels was designated at the position of the minimum radius of the filament.
In addition, because this analysis area depends on the magnification used during shooting, the analysis area must be reduced when decreasing the magnification. 
Similarly, the analysis areas for each magnification were investigated by checking the light intensity images at different magnifications for other solution conditions (Table \ref{Table:camera}).

\section*{References}

\bibliography{aipsamp}

\end{document}